\documentclass[prints]{aa}
\usepackage{graphicx}

\begin{document}
\title{Radial Velocities of Dwarf Spheroidal Galaxies in the M81 Group\thanks{Based on observations collected with the 6m telescope
of the Special Astrophysical Observatory (SAO) of the Russian
Academy of Sciences (RAS), operated under the financial support
of the Science Department of Russia (registration number 01-43).}}
\titlerunning{Dwarf spheroidal galaxies in the M81 group}
\author{M.E.Sharina \inst{1,2} \and I.D.Karachentsev \inst{1}
\and A.N.Burenkov \inst{1,2}}
\authorrunning{M.E.Sharina et al.}
\institute{Special Astrophysical Observatory, Russian Academy
of Sciences, N.Arkhyz, KChR, 369167, Russia,
\and Isaac Newton Institute, Chile, SAO Branch}
\date{Submitted: 30 August 2001}

\abstract{
Long-slit observations of 4 dwarf spheroidal galaxies in the M81 group
are presented. We have obtained a
heliocentric velocity of globular cluster candidate located
near the center of DDO78 to be  $55\ \pm 10\  \mathrm {km}\ \mathrm{s}^{-1}$ by cross-correlation
with template stars.
 We estimated a heliocentric
radial velocity of $-116\ \pm 21\ \mathrm {km}\ \mathrm{s}^{-1}$ for an HII region
seen in the K 61. A red diffuse object near the K64 center is found to be a
remote galaxy with a heliocentric velocity of $+46530\ \mathrm {km}\ \mathrm{s}^{-1}$.
}
\maketitle

\section{Introduction}
 This work continues our study of dwarf spheroidal galaxies (dSphs)
in the M81 group.
 dSphs are diffuse spheroidal objects with central surface brightnesses of $\mu_\mathrm{v}(0) \ga 22 ^m/\sq\arcsec$,
absolute magnitudes M$_\mathrm{V} \ga -14^{m}$,
and HI masses $M_\mathrm{HI} \la 10^5 M_{\sun}$ (Grebel 2000).
Being among the faintest galaxies,
dSphs are rather difficult to observe.
 As gas-poor galaxies, they are usually undetectable in the
HI line. In the case of the M81 group an additional
observational difficulty arises, because of the group location in an area
of the sky that is contaminated with dense Galactic HI emission (Appleton
et al. 1993)  and Galactic cirrus (Sandage 1976).

Almost half of the galaxies in the M81 group are dSphs. There are DDO~71, DDO~78
and K 61, K 64 discovered many years ago by van den Bergh (1959) and
Karachentseva (1968), as well as
BK5N, BK6N, KK77, F8D1, FM1, and KKH57 found in more
recent studies (B\"{o}rngen \& Karachentseva 1982,
Karachentsev 1994, Caldwell et al. 1998, and Froebrich \& Meusinger 2000).
   Based on the presence of HII region, some blue stars and
a probable presence of neutral gas Karachentsev et al.(2000) assumed
K61 to be dSph/dIrr transition type.

  All these objects have been resolved into stars with Hubble Space Telescope
Wide Field Planetary Camera 2 and confirmed as members of the M81 group
via a tip of red giant branch stars (Caldwell et al. 1998, Karachentsev et al. 2000, Karachentsev et al. 2001).

The HST images have allowed us to find globular cluster candidates
in five galaxies of the M81 dSphs sample (Karachentsev et al. 2000).
 The images of four dSph galaxies in the M81 group are presented in Fig.1.
Globular cluster candidates and central background galaxy in the case of K64 are indicated by arrows.
 The images were obtained aboard the Hubble Space Telescope
as part of a snapshot survey of nearby dwarf galaxy candidates (program GO 8192, PI: Seitzer)
and produced by combining the two 600 s exposures taken through the F606W and F814W filters. The galaxies are centred
in the WFC3 chip.

 Basic photometric properties of globular cluster candidates
from Karachentsev et al.(2000) are listed in Table 1.
Its lines contain : (1) integrated apparent magnitude,
(2) integrated color after correction for Galactic reddenning,
(3,4) angular and linear half-light radius, (5) the central
surface brightness, and (6) integrated absolute magnitude.
 Globular clusters in dSphs are relatively bright objects.
In most cases, spectroscopy of them is the only way to get radial velocities of parent galaxies.
Data on radial velocities together with accurate distances for dSphs
allow us to trace the structure and kinematics of the M81 group in more detail.

\section{Observations and data reduction}
The observations were performed with the Long-slit spectrograph
(Afanasiev et al. 1995) at the prime focus of the 6-m SAO (Russia) telescope
during 4 nights in Janury 2001 at a seeing of $\sim 1 \arcsec$ (see Table 2 for
details). The long-slit $130\arcsec$ spectra were obtained with a
CCD-detector having 1024x1024 pixels with 24x24 $ \mu $m
pixel size. For all observations we used the grating of
651 grooves/mm with corresponding dispersion of 2.4 \AA/pixel and a spectral
resolution of 7-9 \AA \hspace{0.5mm}, respectively.
The slit positions (Table 2) were chosen to cross the star cluster candidate
and any characteristic feature of the galaxy, like probable HII regions.
The wavelength range is 4500-6900\AA. In all the cases
the slit width was 2\arcsec \hspace{0.5mm}. The scale along the slit
was 0.39\arcsec/pixel. The reference spectra of Ar-Ne-He lamp
were recorded before and after each observation to provide
wavelength calibration.
For velocity calibration we have obtained long-slit spectra of five
radial velocity standard stars (Barbier-Brossat M., Figon P., 2000)
(see Table 2).
 The spectrophotometric standard star BD28 (Hamuy et al., 1994)
was observed for flux calibration.

  The data reduction was performed using the LONG package
in MIDAS. The subsequent data analysis was also carried out in MIDAS.
 The primary data reduction included cosmic-ray removal,
bias subtraction and flat-field correction. After wavelength
calibration and sky subtraction, the spectra were corrected for atmospheric
extinction and flux-calibrated.
Then rows of every linearized two-dimensional spectrum were summed
in the spatial direction to yield a final one-dimensional spectrum.
All individual exposures of the same object were then co-added to increase
the signal-to-noise ratio. At last, the
spectra of globular cluster candidates and template stars were
divided by normalized continuum and wavelength rebinning was done linearly
in $ \ln \lambda $, as appropriate for the cross-correlation analysis.

\section{Radial velocities}
 Heliocentric radial velocities of the observed objects are summarized in Table 4.
\subsection{DDO78}
DDO78 has a very flat surface brightness profile with the central
surface brightness $\mu_\mathrm{v}(0)=24.5^m/\sq\arcsec$, the exponential scale length $h=28\arcsec$ and
the integrated magnitude $V_\mathrm{t}=15.1$ (Karachentsev et al. 2000).
It is surprising that the galaxy contains the bright globular cluster candidate.
The quality of the DDO78 globular cluster candidate spectrum (Fig.2) is adequate
to derive its radial velocity. The spectra obtained independently in two
observational nights have typically a signal-to-noise ratio of about 10 each.

The spectrum of the globular cluster in DDO 78 was cross-correlated with stellar
template spectra  using the method of Tonry \& Davis (1979).
A procedure of MIDAS "XCORRELATE/IMAGE" correlates two one-dimensional frames
in pixel coordinates.  The spectral resolution is approximately as twice
as the expected internal velocity dispersion in the cluster, so
we have not artificially broadened the template star spectra
before the procedure of cross-correlation.
The globular cluster spectrum was individually cross-correlated
against the velocity templates observed in the corresponding night.
Then the spectrum was divided into four parts.
 These parts were similarly cross-correlated
against the corresponding spectral regions of the velocity templates.
We also estimated the radial velocity by measuring redshifts of
individual absorption lines. The latest method gave a larger rms error
of radial velocity estimates when  compared with the cross-correlation
method. So the mean radial velocity from cross-correlation
served as a final guess to the mean wavelength-shift determination.
The results and errors are summarized in Table 3.

Finaly, we conclude that the observed object is a true
globular cluster belonging to DDO78.
Its mean heliocentric velocity is $+55\ \pm 10\ \mathrm {km}\ \mathrm{s}^{-1}$, where
the error includes "internal" cross-correlation errors and
statistical errors from different cross-correlation manners.

\subsection{K61 and DDO71}
The DDO71 and K61 globular cluster candidates have fainter apparent
magnitudes: $V_\mathrm {T}=20.95$ and $20.70$, respectively (Karachentsev et al. 2000).
Being divided by normalized continuum, their spectra
are presented in Fig.3. The signal-to-noise ratio for them
is poor and the cross-correlation peak vanishes into the noise.
In the case of DDO71 an additional difficulty arises, because
the slit position was chosen to cross
a foreground star. As a result, night sky lines have not been
subtracted correctly.

We have also obtained a spectrum of the HII region in K61 (Fig.4).
Note, that K61 is the brightest dSph galaxy in the M81 group and the closest
companion to M81.
Johnson et al. (1997) revealed a bright HII knot situated NE of the galaxy
center. It shows high excitation emissions with a radial velocity
of  $-135\ \pm 30\ \mathrm {km}\ \mathrm{s}^{-1}$.

Our velocity estimate agrees well with the previous data. The strongest
lines in the spectrum, H$\beta$, [OIII] $\lambda \lambda$ 4959,5007 and
H$ \alpha $, were used to measure the radial velocity of K61. As a result,
the mean heliocentric velocity is $-116\ \pm 21$ $ \mathrm {km}\ \mathrm{s}^{-1}$, where the
error is the standard deviation calculated from different lines.

\subsection{K64}
Binggeli \& Prugniel (1994) ascribed the central star-like object
of K64 as a "quasi-stellar nucleus". However, the large-scale HST images
indicate this object to be a remote red galaxy (Karachentsev et al., 2000).
Our spectral data confirm this conclusion.
Fig.5 shows location of the measured cross-correlation peak.
The spectra of the object and a template star are presented too. We have derived
for the galaxy the  mean heliocentric velocity $46530\  \mathrm {km}\ \mathrm{s}^{-1}$.

\acknowledgements{
 The authors are grateful to A.V.Moiseev, and D.I.Makarov, who
help us to master the methods of Long-slit reduction and cross-correlation.
 We thank A.G.Pramsky for help in observations.

 This work
has been partially supported by the DFG--RFBR grant 01---02---04006.}

\begin{table}
\caption{Properties of globular cluster candidates from Karachentsev et al.(2000).}
\begin{tabular}{lcccc} \\ \hline \hline
Parameter             &DDO78 &  K61 &  DDO71  & K64    \\
\hline
$V_\mathrm{T}$                 & 19.45 & 20.70& 20.95  & (19.5)  \\
$(V-I)_\mathrm{0}$             &  1.07 & 1.00 & 0.99   & (1.44)  \\
$R(0.5L),\;(\arcsec)$ & 0.30  &  0.20& 0.26   &  (1.08) \\
$R(0.5L), \;$ pc      &   5.3 &  3.6 & 4.6    & ---      \\
$\mu_\mathrm{v}(0)$            &  18.0 & 18.5 & 19.2   &  (19.6)   \\
 $\mathrm{M}_\mathrm{v}$                & $-$8.48 & $-$7.37& $-$7.17& ---   \\
\hline
\hline

\multicolumn{5}{l}{K64: The central object appears to be a background galaxy.}
\end{tabular}
\end{table}
\begin{table}
\caption{Observing log }
\scriptsize
\begin{tabular}{llll} \\  \hline \hline
Object               & Date         & Exposure & PA of the slit \\ \hline
Globular cluster     &   18.01.2001 &  2 x 1200 s   & $ 49.^o0 $ \\
candidate            & 18.01.2001   & 2 x 2400  &     \\
in DDO78             &   19.01.2001 &  8 x 1200  &  \\
\\
HII region           & 18.01.2001 & 2 x 1200 & $ 60.^o9  $ \\
and globular cluster &     &       &  \\
candidate in K61     &         &       &  \\
\\
Background galaxy    & 23.01.2001 & 3 x 1800 & $ 60.^o0 $ \\
projected onto K64   &            &         \\
\\
Globular cluster     &   23.01.2001 &  2 x 1800  & $ 78.^o2 $ \\
candidate in DDO71   &            &  2 x 1200  & \\
\\
BF 10078 (F8 V) & 18.01.2001  &  3 x 10  & \\
BF 49601 (G8 V) & 18.01.2001  &  3 x 30  & \\
BF 13987 (F8 VI) & 18.01.2001 &  3 x 60  & \\
BF 18804 (G9 V) & 19.01.2001  &  2 x 20  & \\
BF 18757 (G9 V) & 19.01.2001  &  2 x 10  & \\
\\
BD 28  & 18.01.2001 &  2 x 90  & \\
       & 19.01.2001 &  2 x 150  & \\
\hline  \hline
\end{tabular}
\end{table}

\begin{table}
\caption{Measured heliocentric radial velocities of globular cluster (GC) in DDO78.}
\begin{tabular}{lll} \\  \hline \hline
Method    & $ \lambda \lambda $ \AA  &  $ V_\mathrm{h} $,  $ \mathrm {km}\ \mathrm{s}^{-1}$ \\
\hline  \hline
\sl { Cross-correlation} &                    &     \\
\hline
GC versus BF13987 & $ 4700 \div 6700 $ & +52 \\
		  & $ 4700 \div 5200 $ & +56 \\
		  & $ 6200 \div 6700 $ & +58 \\
GC versus BF18804 & $ 4700 \div 6700 $ & +61 \\
		  & $ 4700 \div 5200 $ & +81 \\
		  & $ 6200 \div 6700 $ & +46 \\
GC versus BF49601 & $ 4700 \div 6700 $ & +40 \\
\hline \hline
\sl { Individual Wavelength Shifts} &       &      \\
\hline
GC regarding to BF18804 & H$ \alpha $ 6562.78,  & $ +76 \pm 39 $ \\
			& Fe+CaI 5270,         &                \\
			& H$\beta$ 4861.33     &                 \\
GC regarding to BF49601 & H$ \alpha $ 6562.78,  & $ +71 \pm 35 $    \\
			& MgI 5172.7, 5183.6,  &                  \\
			&  H$\beta$ 4861.33    &                   \\
\hline  \hline
\end{tabular}
\end{table}
\begin{table}
\caption{Heliocentric radial velocities of the observed objects.}
\begin{tabular}{lc} \\ \hline \hline
Object               & $V_\mathrm{h}$, $ \mathrm {km}\ \mathrm{s}^{-1}$ \\
\hline
Globular cluster     & $55 \pm 10$              \\
candidate in DDO78   &                          \\
\\
HII region in K61    & $116 \pm 21 $ \\
\\
Background galaxy    &  $ 46530 \pm 23 $ \\
projected onto K64   &                   \\
\hline
\hline
\end{tabular}
\end{table}

\onecolumn
\begin{figure}[hbt]
\vbox{\includegraphics{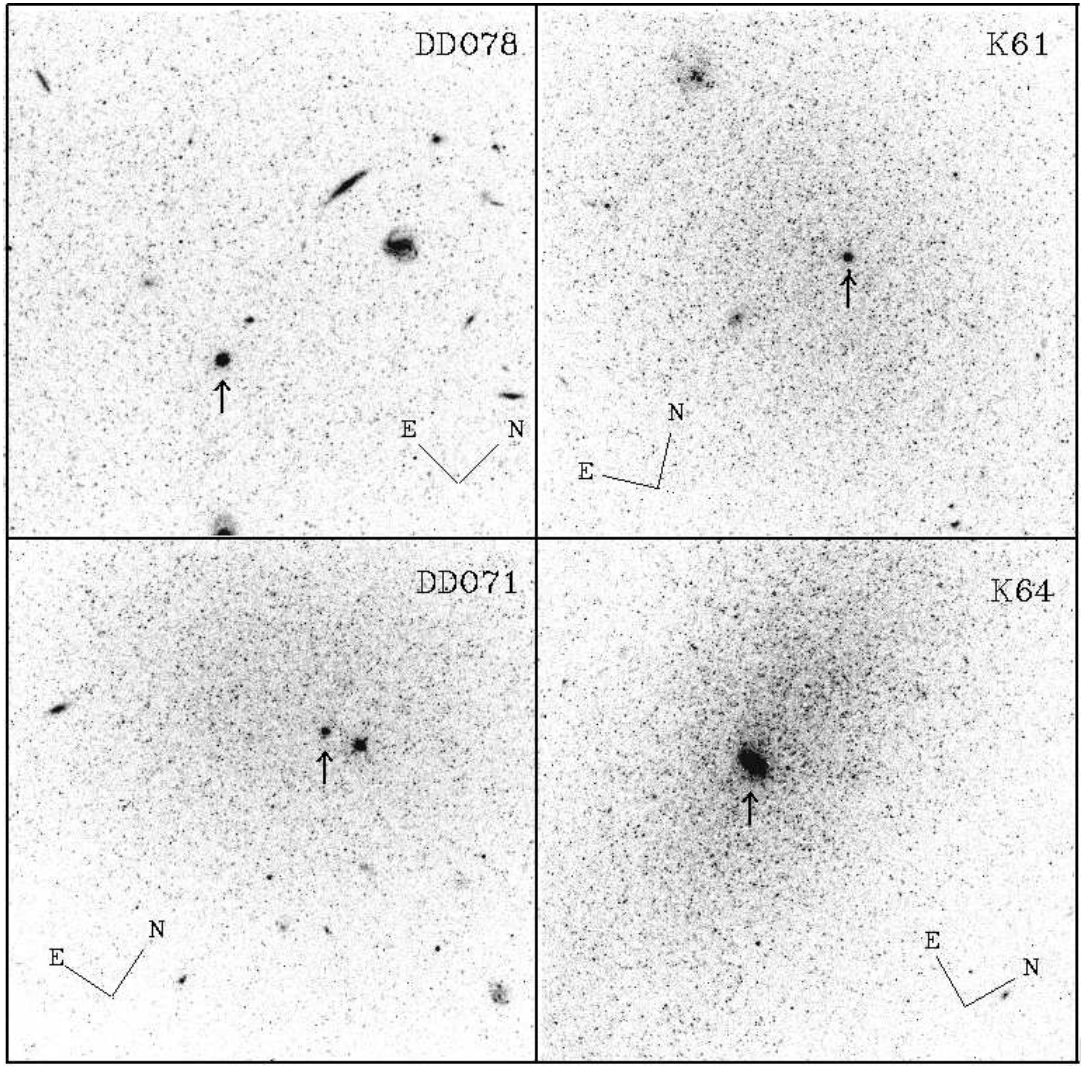}}\par
\vspace{17cm}
\caption{WFPC2 images of four dSph galaxies in the M81 group.
Each galaxy is centered in the WF3 chip (WF3-FIX mode).
The field size of each image is $ 1.3 \arcmin$.
Globular cluster candidates and the central background galaxy in the case of K64 are indicated by arrows.}
\end{figure}
\begin{figure}[hbt]
\vbox{\includegraphics{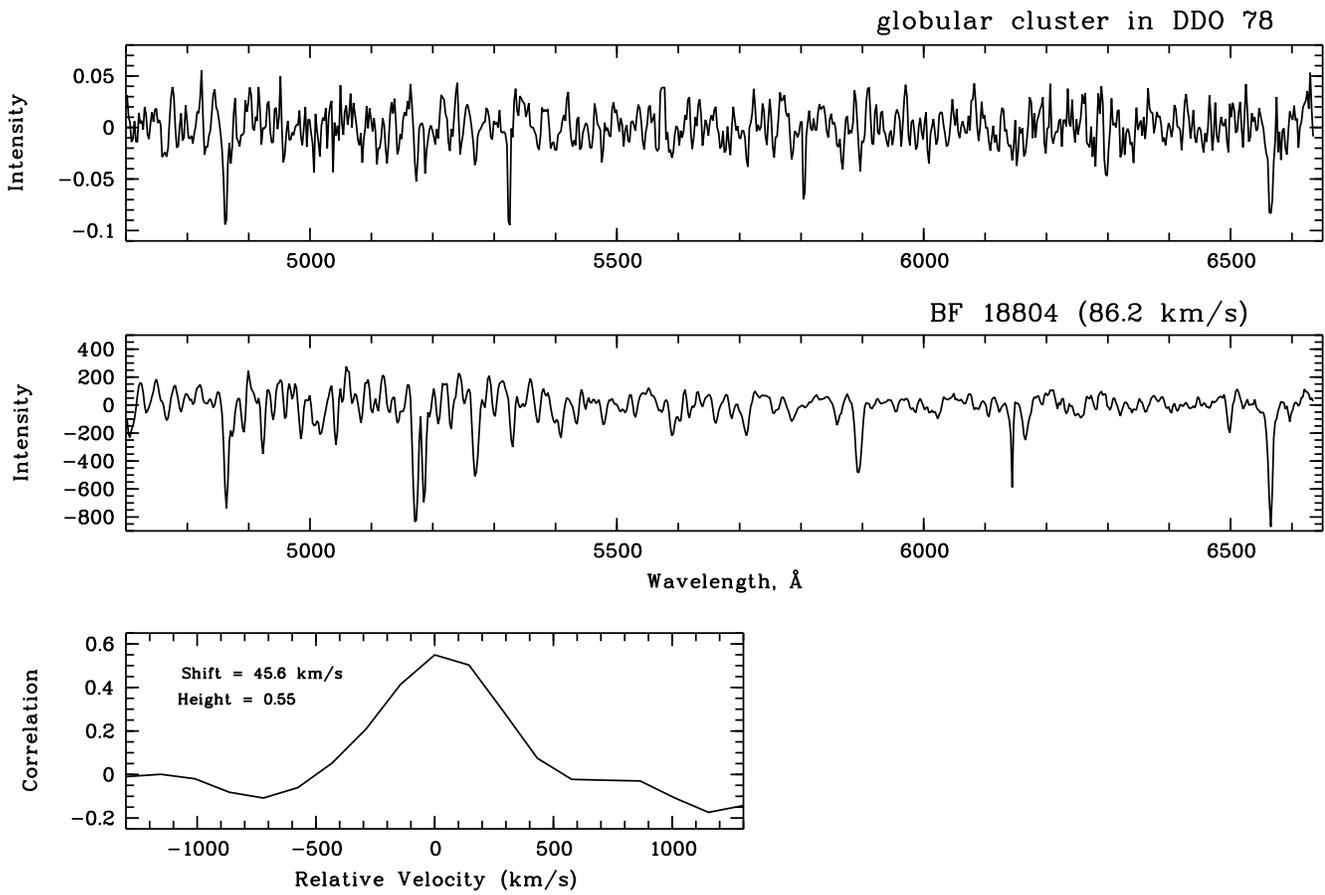}}\par
\vspace{17cm}
\caption{
Cross-correlation function for the
globular cluster candidate in DDO78 vs BF18804.}
\end{figure}
\begin{figure}[hbt]
\vbox{\includegraphics{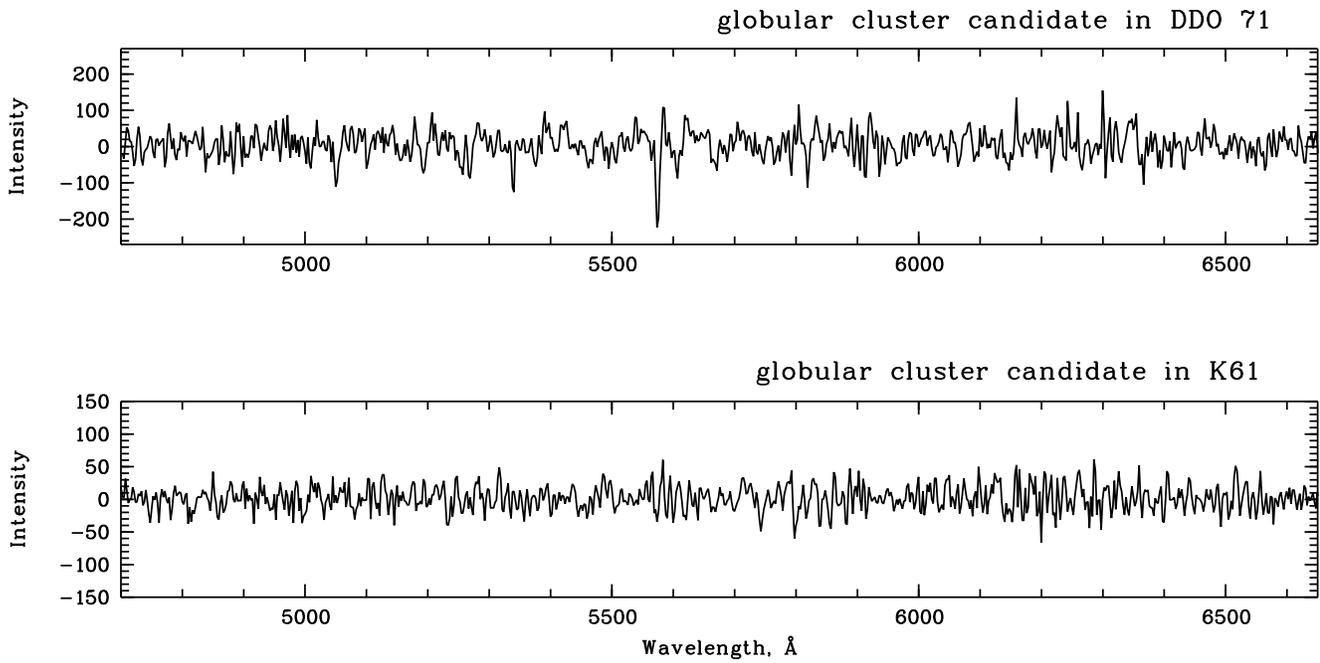}}\par
\vspace{17cm}
\caption{Spectra of globular cluster
candidates in K61 and DDO71 divided by normalized continuum.}
\end{figure}
\begin{figure}[hbt]
\vbox{\includegraphics{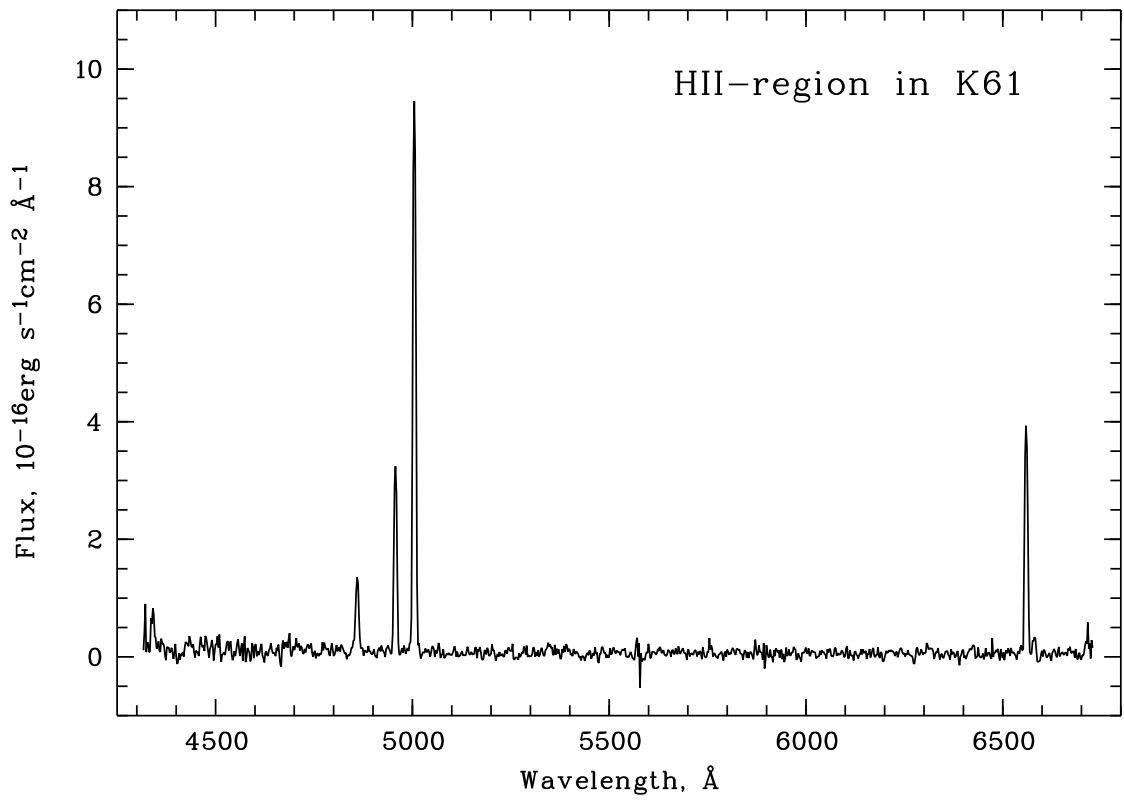}}\par
\vspace{17cm}
\caption{Spectrum of HII region in K61.
}
\end{figure}
\begin{figure}[hbt]
\vbox{\includegraphics{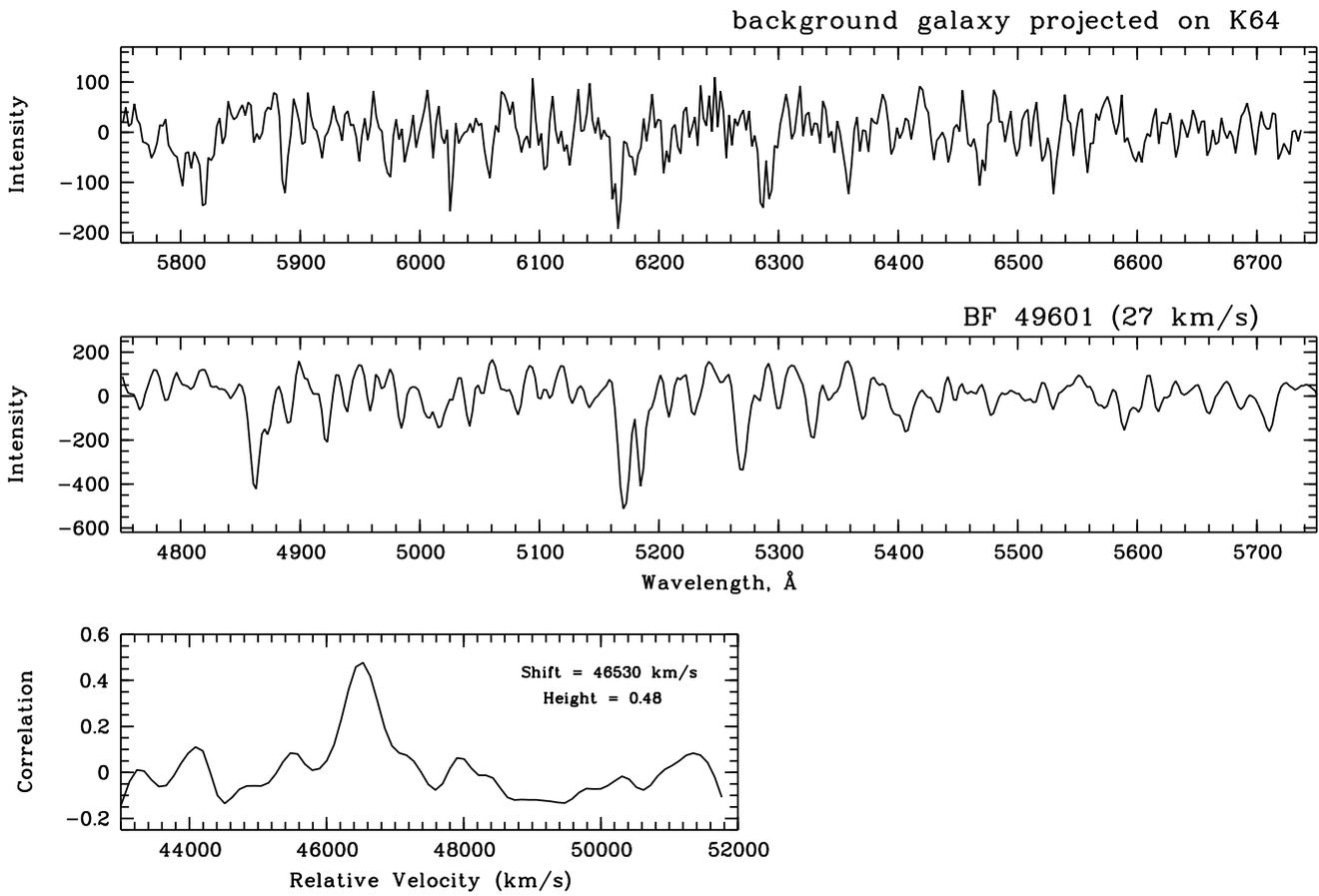}}\par
\vspace{17cm}
\caption{Cross-correlation function for the remote red galaxy
projected onto K64 vs BF49601.}
\end{figure}
\end{document}